\documentclass[aps]{revtex4}
\usepackage{amsfonts}
\usepackage{amsmath}
\usepackage{amssymb,epsf}

\begin{document}

\title{$(2+1)$-dimensional solutions in $F(R)$ gravity}
\author{S. H. Hendi\footnote{E-mail: hendi@shirazu.ac.ir}}
\affiliation{Physics Department and Biruni Observatory, College of Sciences, Shiraz
University, Shiraz 71454, Iran \\
Research Institute for Astrophysics and Astronomy of Maragha (RIAAM), P.O.
Box 55134-441, Maragha, Iran}

\begin{abstract}
Motivated by the well-known charged BTZ black holes, we look for
$(2+1)$-dimensional solutions of $F(R)$ gravity. At first we
investigate some near horizon solutions and after that we obtain
asymptotically Lifshitz black hole solutions. Finally, we discuss
about rotating black holes with exponential form of $F(R)$ theory.
\end{abstract}

\maketitle

\section{Introduction}

Regardless of the dark energy nature, one of the most popular ad hoc
theories to explain the current accelerated expansion of our universe is
dark energy \cite{DE}. Necessity of dark energy comes from the fact that we
need to have a negative pressure (repulsive action) to interpret cosmic
expansion. Since the determination of dark energy nature is an important
challenge for the physics communities, it is inevitable to look for an
alternative theories scenario for dark energy to address the observational
evidences. Modified gravity theory, instead of general relativity, is an
alternative plan to describe the late time acceleration \cite{ModGrav}.

In recent years, variety of Modified theories of classical gravity
have been proposed to solve some puzzles of standard general
relativity. Amongst them the well-known $F(R)$ theory, whose
Lagrangian density is an arbitrary function of the Ricci scalar,
is quite special and received a growing attention (see for example
\cite{HendiGRG} and references therein). $F(R)$ gravity provides a
technically powerful tool to deal with the early time inflation
\cite{Early}, late time acceleration \cite{Late}, the hierarchy
and singularity problems \cite{hierarchy,singularity} and (the
nature of) dark energy \cite{F(R)DE}. Holographic superconductor
with linear and nonlinear Maxwell field in the frame of modified
gravity has been studied \cite{F(R)Max,F(R)nonMax} and the
condensation effects of nonlinearity in Maxwell field and
curvature terms have been investigated in \cite{F(R)nonMax}.
Although, the field equations of $F(R)$ theories are of four-order
and solving them, directly, is so complicated, their valuable
consequences motivate us to consider them and investigate their
interesting properties. Using a suitable conformal transformation,
it has been shown that $F(R)$ gravity models are equivalent to
classical Einstein's gravity with an extra scalar field. Also, we
can apply some limitations on the model parameters to guarantee
that the model follows the stability condition (the scalaron is
not a tachyon) and has no ghosts \cite{ghosts,Nojiri-Odintsov}.
Some viable models of $F(R)$ theories have been widely
investigated in the literature
over the past few years \cite%
{F(R)1,Hu-Sawicki,Starobinsky,Dev,Appleby-Battye,Nojiri-Odintsov,Tsujikawa}.

In addition to the $F(R)$ theories, one of the interesting subjects for
recent study is the investigation of three dimensional black holes \cite%
{BTZnew}. Considering three dimensional spacetimes helps us to find some
conceptual issues in the black hole properties, quantum view of gravity and
string theory \cite{BTZ1,BTZ2}. Therefore, theoretical physicists have an
interest in the $(2+1)$-dimensional manifolds and their properties \cite%
{BTZ3}. Moreover, three dimensional solutions perform an essential role to
improve our comprehension of gravitational interaction in low dimensional
manifolds \cite{BTZ4}. In addition, it is interesting to study the
asymptotic behavior as well as near horizon solutions of BTZ black holes
\cite{BTZ6} and generalize its properties to higher dimensions \cite{BTZ5}.
In this work, we investigate some interesting solutions of $F(R)$ gravity in
$(2+1)$-dimensions.

The organization of the paper is as follows: at first, we give a brief
review of the field equations of $F(R)$ gravity. In the next section, we
obtain the near horizon solution for $F(R)$ gravity in three dimensional
static spacetime. After that, we look for the existence of exact solutions
of some interesting models. We finish our paper with some conclusions.

\section{ Basic Field Equations and metric ansatz \label{FieldF(R)}}

Action of $F(R)$ gravity in the presence of matter field in
arbitrary dimensions and its related field equations have been
obtained before \cite{Dombriz}. In addition, static and
spherically symmetric solutions of F(R)
gravity with constant Ricci scalar have been investigated \cite%
{HendiGRG,Dombriz}. Following Refs. \cite{HendiGRG,Dombriz}, one finds that
the action of $3$-dimensional $F(R)=R+f(R)$ gravity, in the presence of
matter field has the form of
\begin{equation}
\mathcal{I}=-\frac{1}{16\pi }\int d^{3}x\sqrt{-g}[R+f(R)]+\mathcal{I}_{matt},
\label{Action}
\end{equation}%
where $\mathcal{I}_{matt}$ is the action of matter fields. One can vary Eq. (%
\ref{Action}) with respect to the metric $g_{\mu \nu }$ to obtain the
following field equations
\begin{equation}
R_{\mu \nu }(1+f_{R})-\frac{1}{2}g_{\mu \nu }\left[ R+f(R)\right] +(\nabla
_{\mu }\nabla _{\nu }-g_{\mu \nu }\nabla _{\beta }\nabla ^{\beta })f_{R}=%
\mathrm{T}_{\mu \nu }^{matt},  \label{FE}
\end{equation}%
where $f_{R}\equiv df(R)/dR$ and $\mathrm{T}_{\mu \nu }^{matt}$ is the
standard matter stress-energy tensor which is derived from the matter action
$\mathcal{I}_{matt}$ in Eq. (\ref{Action}). Here, we want to obtain the $3$%
-dimensional static and spherically symmetric solutions of Eq.
(\ref{FE}). We assume the metric has the following ansatz
\begin{equation}
ds^{2}=-g(r)dt^{2}+\frac{dr^{2}}{g(r)}+r^{2}d\phi ^{2}.  \label{Met1}
\end{equation}%
Considering the sourceless ($\mathrm{T}_{\mu \nu }^{matt}=0$) field equation
(\ref{FE}) with the metric (\ref{Met1}), one can obtain the following
independent differential equations
\begin{eqnarray}
&&2rg(r)f_{R}^{\prime \prime }+\left[ rg^{\prime }(r)+2g(r)\right]
f_{R}^{\prime }+\left[ rg^{\prime \prime }(r)+g^{\prime }(r)\right]
f_{R}-g^{\prime }(r)=-rf(R),  \label{Fieldtt} \\
&&\left[ rg^{\prime }(r)+2g(r)\right] f_{R}^{\prime }+\left[ rg^{\prime
\prime }(r)+g^{\prime }(r)\right] f_{R}-g^{\prime }(r)=-rf(R),
\label{Fieldrr} \\
&&2rg(r)f_{R}^{\prime \prime }+2rg^{\prime }(r)f_{R}^{\prime }+2g^{\prime
}(r)f_{R}-rg^{\prime \prime }(r)=-rf(R),  \label{Fieldthth}
\end{eqnarray}%
corresponding to $tt$, $rr$ and $\phi \phi $ components of Eq. (\ref{FE}),
respectively. It is notable that the prime and double prime are,
respectively, the first and second derivatives with respect to $r$. Here, we
study black hole solutions with constant Ricci scalar (so $f_{R}^{\prime
\prime }=f_{R}^{\prime }=0$), and therefore it is easy to show that the
field equations (\ref{Fieldtt})-(\ref{Fieldthth}) reduce to
\begin{eqnarray}
\left[ rg^{\prime \prime }(r)+g^{\prime }(r)\right] f_{R}-g^{\prime }(r)
&=&-rf(R),  \label{eqaa} \\
2g^{\prime }(r)f_{R}-rg^{\prime \prime }(r) &=&-rf(R),  \label{eqbb}
\end{eqnarray}%
Equating the left hand sides of Eqs. (\ref{eqaa}) and (\ref{eqbb}), we
obtain
\begin{equation}
\left[ rg^{\prime \prime }(r)-g^{\prime }(r)\right] \left( f_{R}+1\right) =0,
\label{eqc}
\end{equation}%
with the trivial uncharged static BTZ solution $g(r)=\frac{r^{2}}{l^{2}}-M$
with arbitrary $f(R)$, and also $f(R)=-R$ for arbitrary $g(r)$. It is
important to note that we are looking for a solution which satisfies both
Eqs. (\ref{eqaa}) and (\ref{eqbb}), simultaneously, and the mentioned
trivial BTZ solution is the solution of them for arbitrary $f(R)$. In
addition, considering $f(R)=-R$ ($F(R)=0$), both Eqs. (\ref{eqaa}) and (\ref%
{eqbb}) are satisfied for $g(r)=\frac{R_{0}}{6}r^{2}-M+\frac{C}{r}$, not
arbitrary $g(r)$. As we will see, we can interpret $C$ as an electric charge
parameter and so one can obtain charged solution of $F(R)$ gravity for the
special case $f(R)=-R$. There are some interesting notes arising from this
solution. We find that the field equations of $F(R)$ action admit a static
solution as $g(r)=\frac{R_{0}}{6}r^{2}-M+\frac{C}{r}$ for $f(R)=-R$ . On the
other hand, one can obtain the same metric function $g(r)=-\Lambda r^{2}-M+%
\frac{q}{r}$ for charged black hole Einstein-power Maxwell invariant
(Einstein-PMI) gravity \cite{HendinonlinearEPJC} when the nonlinearity
parameter is chosen $s=3/4$ with following action
\[
\mathcal{I}=-\frac{1}{16\pi }\int d^{3}x\sqrt{-g}[R-2\Lambda +(-F_{\mu \nu
}F^{\mu \nu })^{s}].
\]%
This result implies two interesting results. First, one can see the the
effects of the cosmological constant and charge term of the PMI metric can
be reproduced by $f(R)=-R$ in the $F(R)$ gravity. So one may like to
interpret it as a kind of link between gravitational theory and a classical
field theory like PMI. Indeed, $F(R)$ gravity provides a framework for
putting the gravity and nonlinear electrodynamics in a unified context by
using geometry. In other words, one may ask: in geometric point of view, can
we consider PMI Lagrangian $(-F_{\mu \nu }F^{\mu \nu })^{s}$ equivalent to $%
(2\Lambda -R)$?

Second, from $F(R)$ gravity point of view of this solution, action become
zero and there is not any well-defined thermodynamic potential for this
theory \cite{foot} and one can only talk about the temperature of the black
hole. On the other hand, the metric of the Einstein-PMI theory is described
by well-defined thermodynamic properties like entropy and so on. By this
observation, one may ask: what is the role of the geometry of the spacetime
in the black hole thermodynamics? Is it possible to define an entropy
corresponding to the horizon of the black hole in the context of $F(R)$
gravity or not?

\section{Black hole solutions of the $F(R)=R+f(R)$ gravity \label{SolF(R)}}

\subsection{Near horizon solution}

The idea of studying the near horizon black hole has great appeal and a long
history \cite{near}. Considering some of viable complicated theories of
gravity, one cannot find an (easy) exact solution. Therefore, we try to
obtain approximate or numerical solutions with suitable boundary conditions
\cite{HendiMomeni}. Now, we consider near horizon solutions for some models
of $F(R)$ gravity in three dimensional static spacetime.

\subsubsection{CaseI: $f(R)=-2\Lambda $}

To demonstrate this method, here, we consider a trivial well-known case $%
f(R)=-2\Lambda$, with constant curvature i.e. $R=R_{0}$. Thus the equation
of motion (\ref{FE}) reduces to
\begin{equation}
R_{\mu \nu }=g_{\mu \nu }\left( \frac{1}{2}R_{0}-\Lambda \right) ,
\label{RicciEq}
\end{equation}
Applying the metric (\ref{Met1}) to Eq. (\ref{RicciEq}), one may obtain
\begin{eqnarray}
g^{\prime }(r) &=&-2\Lambda r,  \label{dg} \\
g^{\prime \prime }(r) &=&-2\Lambda .  \label{ddg}
\end{eqnarray}
Here, we would like to obtain the near horizon solution and compare it with
the exact one. It is easy to show that the exact solution of Eqs. (\ref{dg})
and (\ref{ddg}) is
\begin{equation}
g(r)=-\Lambda r^{2}-M,  \label{exact}
\end{equation}
but the procedure is different for the near horizon black hole solution.
According to the Hawking-Bekenstein temperature formula, if the metric (\ref%
{Met1}) posses a black hole solution with an event horizon located at $%
r=r_{+}$, we can deduce $g(r_{+})=0$ and
\begin{equation}
T=\frac{g^{\prime }(r_{+})}{4\pi }.  \label{T3}
\end{equation}
Expanding the metric function $g(r)$ near the horizon, one can obtain
\begin{equation}
g(r)=\frac{(r-r_{+})}{1!}g^{\prime }(r_{+})+\frac{(r-r_{+})^{2}}{2!}
g^{\prime \prime }(r_{+})+\frac{(r-r_{+})^{3}}{3!}g^{\prime \prime \prime
}(r_{+})+\frac{(r-r_{+})^{4}}{4!}g^{\prime \prime \prime \prime }(r_{+})+...,
\label{Bast}
\end{equation}
where in this $f(R)$ model, the nonvanishing terms of Eq. (\ref{Bast}) are
the first two terms. Thus, the near horizon solution of $g(r)$ is obtained
as
\begin{eqnarray}
g(r) &=&4\pi T(r-r_{+})-(r-r_{+})^{2}\Lambda  \nonumber \\
&=&-r^{2}\Lambda +2\left( 2\pi T+\Lambda r_{+}\right) r-r_{+}^{2}\Lambda
-4\pi Tr_{+}.  \label{gnear1}
\end{eqnarray}
Considering both Eqs. (\ref{dg}) and (\ref{T3}), one can show that $2 \pi
T=-\Lambda r_{+}$ and therefore Eq. (\ref{gnear1}) reduces to
\begin{equation}
g(r)=-\Lambda r^{2}+\Lambda r_{+}^{2}.  \label{gnear2}
\end{equation}
In order to obtain the exact solution (\ref{exact}) from the near horizon
solution, it is sufficient to set $M=-\Lambda r_{+}^{2}$ in Eq. (\ref{gnear2}%
). This adjustment may also come from Eq. (\ref{exact}), in which $%
g(r_{+})=0 $. Hence, for the mentioned specific trivial model, the near
horizon solution is matched to exact solution.

\subsubsection{Case II: arbitrary $f(R)$ with constant $R$:}

In this subsection, we apply the recent procedure to an arbitrary model of $%
f(R)$ with constant Ricci scalar. We can rewrite Eqs. (\ref{eqaa}) and (\ref%
{eqbb}) in the following forms
\begin{eqnarray}
g^{\prime \prime }(r) &=&\frac{1}{rf_{R}}\left[ \left( 1-f_{R}\right)
g^{\prime }(r)-rf\right] ,  \label{ddga} \\
g^{\prime \prime }(r) &=&f+2f_{R}\frac{g^{\prime }(r)}{r}.  \label{ddgb}
\end{eqnarray}%
Equating the right hand side of both Eqs. (\ref{ddga}) and (\ref{ddgb}), one
may obtain two solutions for $f_{R}$
\begin{equation}
f_{R}=-1,\frac{1}{2}-\frac{rf}{2g^{\prime }(r)},  \label{f_R}
\end{equation}%
where we are not interested in the first solution ($f_{R}=-1$),
here. One can use the second solution
($f_{R}=\frac{1}{2}-\frac{rf}{2g^{\prime }(r)}$) and the
definition of the black hole temperature $g^{\prime }(r_{+})=4\pi
T$ in Eq. (15) to obtain the near horizon solution
\begin{equation}
g(r)=2\pi T\frac{(r^{2}-r_{+}^{2})}{r_{+}}=\frac{2\pi T}{r_{+}}r^{2}-2\pi
Tr_{+}.
\end{equation}%
It is easy to set $\Lambda =-2\pi T/r_{+}$ and $M=2\pi Tr_{+}$ to obtain
three dimensional solution of Einstein-$\Lambda $ gravity. Therefore, the
near horizon solution of arbitrary $F(R)$ gravity models with constant $R$
is the same as uncharged BTZ solution.

\subsubsection{Case III: arbitrary $f(R)$ with nonconstant $R$:}

Here, we take into account an arbitrary model of $f(R)$ with nonconstant
Ricci scalar. One can consider Eqs. (\ref{Fieldtt})-(\ref{Fieldthth}) and
solve them near the horizon to obtain
\begin{eqnarray}
g(r_{+}) &=&0,  \label{gh} \\
g^{\prime }(r_{+}) &=&\frac{f_{+}\left[ 1+f_{R+}\right] r_{+}}{%
1-f_{R+}-f_{R+}^{\prime }r_{+}-2f_{R+}^{2}-2f_{R+}f_{R+}^{\prime }r_{+}},
\label{dgh} \\
g^{\prime \prime }(r_{+}) &=&\frac{f_{+}\left[ 1+f_{R+}+f_{R+}^{\prime }r_{+}%
\right] }{1-f_{R+}-f_{R+}^{\prime }r_{+}-2f_{R+}^{2}-2f_{R+}f_{R+}^{\prime
}r_{+}},  \label{ddgh}
\end{eqnarray}%
where $f_{+}=\left. f(R)\right\vert _{r=r_{+}}$, $f_{R+}=\left.
f_{R}\right\vert _{r=r_{+}}$ and $f_{R+}^{\prime }=\left. \frac{df_{R}}{dr}%
\right\vert _{r=r_{+}}$. Using the fact that $g^{\prime }(r_{+})=4\pi T$ \
with Eq. (\ref{dgh}), we achieve%
\begin{eqnarray}
f_{+} &=&\frac{4\pi T\left[ 1-f_{R+}-f_{R+}^{\prime
}r_{+}-2f_{R+}^{2}-2f_{R+}f_{R+}^{\prime }r_{+}\right] }{\left[ 1+f_{R+}%
\right] r_{+}},  \label{fp} \\
g^{\prime \prime }(r_{+}) &=&\frac{4\pi T\left[ 1+f_{R+}+f_{R+}^{\prime
}r_{+}\right] }{\left[ 1+f_{R+}\right] r_{+}},  \label{ddgh2}
\end{eqnarray}%
and therefore the near horizon solution (Eq. (\ref{Bast}) up to second
order) reduces to%
\begin{eqnarray}
g(r) &=&4\pi T\left( r-r_{+}\right) +\frac{2\pi T\left( r-r_{+}\right) ^{2}%
\left[ 1+f_{R+}+f_{R+}^{\prime }r_{+}\right] }{\left( 1+f_{R+}\right) r_{+}}
\nonumber \\
&=&\frac{2\pi T\left[ 1+f_{R+}+f_{R+}^{\prime }r_{+}\right] }{r_{+}\left(
1+f_{R+}\right) }r^{2}-\frac{4\pi Tf_{R+}^{\prime }r_{+}}{1+f_{R+}}r-\frac{%
2\pi T\left[ 1+f_{R+}-f_{R+}^{\prime }r_{+}\right] r_{+}}{1+f_{R+}}.
\label{gh2}
\end{eqnarray}%
Eq. (\ref{gh2}) is a second order function such as BTZ solution
with additional linear term. In other word, we can compare Eq.
(\ref{gh2}) and BTZ solution with the following equalities
\begin{eqnarray}
\frac{2\pi T\left[ 1+f_{R+}+f_{R+}^{\prime }r_{+}\right] }{r_{+}\left(
1+f_{R+}\right) } &=&-\Lambda ,  \label{Lambda} \\
\frac{2\pi T\left[ 1+f_{R+}-f_{R+}^{\prime }r_{+}\right] r_{+}}{1+f_{R+}}
&=&M.  \label{M}
\end{eqnarray}%
After straightforward calculations, one can use Eqs. (\ref{Lambda}) and (\ref%
{M}) to achieve%
\begin{eqnarray}
f_{R+}^{\prime } &=&\frac{\left( \Lambda r_{+}^{2}+M\right) \left(
1+f_{R+}\right) }{\left( \Lambda r_{+}^{2}-M\right) r_{+}},  \label{fprim} \\
T &=&-\frac{\left( \Lambda r_{+}^{2}-M\right) }{4\pi r_{+}^{2}}.  \label{T}
\end{eqnarray}
Inserting Eqs. (\ref{fprim}) and (\ref{T}) in Eq. (\ref{gh2}), we obtain%
\begin{eqnarray}
g(r) &=&\left( r-r_{+}\right) \frac{\left( rr_{+}+Ml^{2}\right) }{l^{2}r_{+}}
\nonumber \\
&=&-\Lambda r^{2}+\frac{\left( \Lambda r_{+}^{2}+M\right) }{r_{+}}r-M,
\label{gh3}
\end{eqnarray}%
which is the BTZ solution with additional linear term. This linear term
comes from the fact that we chose a nonconstant Ricci scalar solution. As we
have seen in case II, this linear term vanishes for the solutions with
constant Ricci scalar. As one can confirm, the linear term of Eq. (\ref{gh3}%
) does not change the horizon class and asymptotic behavior of the spacetime
and therefore, not only is it not necessary to remove it but also it will be
interesting to think about its physical interpretation.

\subsection{Hyperscaling violation and Lifshitz exact solutions:}

In order to obtain the exact solutions, we should choose a specific model
for $F(R)$. We should consider $F(R)$ models in which local gravity
constraints are satisfied as well as cosmological and stability conditions.
We know that some of the viable and interesting forms of $F(R)$ gravity have
been considered by Hu-Sawicki \cite{Hu-Sawicki}, Starobinsky \cite%
{Starobinsky} and its generalization \cite{Dev}, Appleby-Battye \cite%
{Appleby-Battye}, Nojiri-Odintsov \cite{Nojiri-Odintsov} and Tsujikawa \cite%
{Tsujikawa}. In what follows, we consider two kinds of these models to
obtain asymptotically Lifshitz solutions. For other models, the method is
straightforward.

\subsubsection{Hu-Sawicki model: $F(R)=R-m^{2}\frac{c_{1}\left( \frac{R}{
m^{2}}\right) ^{n}}{1+c_{2}\left( \frac{R}{m^{2}}\right) ^{n}}$}

In this section, we are looking for the asymptotically Lifshitz solution
with a hyperscaling overall factor for this kind of model. In order to
achieve this goal, we consider the following ansatz
\begin{equation}
ds^{2}=r^{\alpha }\left[ -\left( \frac{r^{2}}{l^{2}}\right) ^{z}g(r)dt^{2}+%
\frac{l^{2}dr^{2}}{r^{2}g(r)}+r^{2}d\phi ^{2}\right] ,  \label{ALifMet}
\end{equation}%
where the constants $z$ and $\alpha $ are called dynamical and hyperscaling
violation exponents, respectively. For simplicity, we can set $\alpha =-2$,
now considering asymptotically Lifshitz metric (\ref{ALifMet}) with the
mentioned $F(R)$ model. One can obtain the metric function $g(r)$ with the
following relation
\begin{equation}
g(r)=\left( a+\frac{b}{r^{z-2}}\right) r^{-z}-\frac{l^{2}R_{0}}{%
2r^{2}(z-2)^{2}}.  \label{gLif}
\end{equation}%
Inserting the mentioned metric in the field equations, we obtain a set of
algebraic equations for the model parameters as
\begin{eqnarray}
(n-1)m^{2n}-c_{2}R_{0}^{n} &=&0, \\
\frac{m^{2}c_{1}R_{0}^{n-1}}{\left[ m^{2n}+c_{2}R_{0}^{n}\right] }-1 &=&0,
\end{eqnarray}%
with the following solutions%
\begin{eqnarray}
c_{1} &=&\frac{nm^{2n-2}}{R_{0}^{n-1}},  \label{c1} \\
c_{2} &=&(n-1)\frac{m^{2n}}{R_{0}^{n}},  \label{c2}
\end{eqnarray}%
where the constant $R_{0}$ is Ricci scalar. We calculate the nonvanishing
components of Riemann tensor for obtained solution and find that
\begin{equation}
R^{trtr}=-\frac{R_{0}l^{2z-2}}{2r^{2z-6}},
\end{equation}%
which confirm that for $z\neq 3$, there is an essential singularity at the
origin ($r=0$).

It is notable that existence of $r^{-2}$ factor in the last term of Eq. (\ref%
{gLif}) excludes any Lifshitz solution with hyperscaling factor. In the case
of $\alpha =0$ with the mentioned model parameters, one can obtain an
asymptotically Lifshitz solution%
\begin{eqnarray}
ds^{2} &=&-\left( \frac{r^{2}}{l^{2}}\right) ^{z}g(r)dt^{2}+\frac{l^{2}dr^{2}%
}{r^{2}g(r)}+r^{2}d\phi ^{2},  \label{metLifz} \\
g(r) &=&\left( ar^{\frac{\sqrt{z^{2}+4z-4}}{2}}+\frac{b}{r^{\frac{\sqrt{%
z^{2}+4z-4}}{2}}}\right) r^{-\frac{3z}{2}-1}-\frac{l^{2}R_{0}}{2(z^{2}+z+1)}.
\label{gLifz}
\end{eqnarray}

Moreover, in this situation, we can adjust $a=b=0$ and $%
R_{0}=-2(z^{2}+z+1)/l^{2}$ to find Lifshitz solution as a vacuum solution of
the theory. It is easy to show that the Kretschmann scalar diverges at $%
r\longrightarrow 0$ and for large values of $r$ one obtains
\begin{equation}
\lim_{r\longrightarrow \infty }R_{\alpha \beta \mu \nu }R^{\alpha \beta \mu
\nu }=\frac{z^{2}-z+1}{z^{2}+z+1}R_{0}^{2}.  \label{RiemLif}
\end{equation}%
Here, we want to add angular momentum to static Lifshitz metric to
obtain possible rotating solution. We take into account the
following ansatz
\begin{equation}
ds^{2}=-\left( \frac{r^{2}}{l^{2}}\right) ^{z}g(r)dt^{2}+\frac{l^{2}dr^{2}}{%
r^{2}g(r)}+r^{2}\left[ d\phi +h(r)dt\right] ^{2}.  \label{RotLifshitz}
\end{equation}%
Inserting this rotating metric in the field equations with Eqs. (\ref{c1})
and (\ref{c2}), one can find the following solutions for the metric functions%
\begin{eqnarray}
h(r) &=&Br^{-\eta },  \label{h} \\
g(r) &=&\left( ar^{\frac{\sqrt{z^{2}+4z-4}}{2}}+\frac{b}{r^{\frac{\sqrt{%
z^{2}+4z-4}}{2}}}\right) r^{-\frac{3z}{2}-1}  \nonumber \\
&&+\frac{B^{2}\eta ^{2}l^{2z}r^{2-2\eta -2z}}{4\left[ 2\eta ^{2}-6\eta +\eta
z-2z+5\right] }-\frac{l^{2}R_{0}}{2(z^{2}+z+1)}.  \label{gRL}
\end{eqnarray}%
We should note that for $B=0$, this solution reduces to the static Lifshitz
solution, as it should be. It has been shown that considering the dynamical
field of Ricci scalar, the effective mass is related to $\frac{d^{2}F}{dR^{2}%
}$ \cite{DKstability1,DKstability2}. Therefore, in order to obtain a stable
dynamical field, its effective mass must be positive. This requirement is
known as the Dolgov-Kawasaki stability criterion \cite{DKstability2}. It is
notable that the second derivative of the $F(R)$ function for this specific
model is
\begin{equation}
F_{RR}=\frac{n-1}{R_{0}},  \label{ddF}
\end{equation}%
which is positive for positive $R_{0}$\ and $n>1$.

\subsubsection{Modified Starobinsky model \protect\cite{Dev}: $F(R)=R+%
\protect\lambda \protect\beta \left( \left[ 1+\left( \frac{R}{\protect\beta }%
\right) ^{2}\right] ^{-n}-1\right) +\protect\kappa \frac{R^{2}}{\protect%
\beta }$}

It is easy to show that taking into account the mentioned modified
Starobinsky model with metrics (\ref{ALifMet}), (\ref{metLifz})
and (\ref{RotLifshitz}), one can obtain the same relation for
$g(r)$ as presented in Eqs. (\ref{gLif}), (\ref{gLifz}) and
(\ref{gRL}). We should note that in order to satisfy all field
equations, we should set the model parameters
\begin{eqnarray}
\kappa &=&-\frac{\beta }{2R_{0}}-\frac{n\beta R_{0}}{2(n+1)R_{0}^{2}+2\beta
^{2}\left( 1-\left[ 1+\left( \frac{R_{0}}{\beta }\right) ^{2}\right]
^{n+1}\right) },  \label{kappa} \\
\lambda &=&-\frac{\beta R_{0}}{\frac{2(n+1)R_{0}^{2}+2\beta ^{2}}{\left[
1+\left( \frac{R_{0}}{\beta }\right) ^{2}\right] ^{n+1}}-2\beta ^{2}}.
\label{lambda}
\end{eqnarray}

In order to confirm the Dolgov-Kawasaki stability, we obtain
\begin{eqnarray}
F_{RR} &=&\frac{1-\frac{(2n+1)(n+1)(\Xi -1)^{2}+(n+2)(\Xi -1)+1}{\Xi ^{n+2}}%
}{R_{0}\left( \frac{1+(n+1)(\Xi -1)}{\Xi ^{n+1}}-1\right) }, \\
\Xi &=&1+\left( \frac{R_{0}}{\beta }\right) ^{2}.  \nonumber
\end{eqnarray}%
It is easy to show that for special values of $n$ and $\beta $, we get
positive $F_{RR}$. It should also be noted that one can obtain the same
exact solutions for most of viable models (or their generalizations) by
setting the model parameters, suitably.

\subsubsection{Rotating solution: case I:}

Here, we consider a rotating spacetime as
\begin{equation}
ds^{2}=-g(r)dt^{2}+\frac{dr^{2}}{g(r)}+r^{2}\left( d\phi +\frac{b}{r^{2}}%
dt\right) ^{2}.  \label{Met2}
\end{equation}%
where $b$ is a constant. Regarding Eq. (\ref{Met2}) with the field equations
(\ref{eqaa}) and (\ref{eqbb}), one may obtain
\begin{equation}
g(r)=Ar^{2}-M+\frac{b^{2}}{r^{2}},  \label{g}
\end{equation}%
where $A$ and $M$ are integration constants and%
\begin{equation}
f(R)=\frac{g^{\prime }(r)}{r}+\frac{2b^{2}}{r^{4}}+C\exp \left( \frac{-r^{4}R%
}{4b^{2}+2r^{3}g^{\prime }(r)}\right) .  \label{f(R)rot}
\end{equation}%
Applying Eq. (\ref{g}) in Eq. (\ref{f(R)rot}) and set $A=-\Lambda $, one can
obtain the exponential correction of Einstein gravity for the $f(R)$ model.
In other words, Eq. (\ref{f(R)rot}) reduces to
\begin{equation}
f(R)=-2\Lambda +C\exp \left( \frac{R}{4\Lambda }\right) .
\end{equation}%
In order to analyze the geometric properties of the solution, we calculate
the nonzero components of Riemann tensor
\begin{eqnarray*}
R^{trtr} &=&-\Lambda , \\
R^{trr\phi } &=&\frac{-\Lambda b}{r^{2}}, \\
R^{t\phi t\phi } &=&\frac{\Lambda }{\Lambda r^{4}+Mr^{2}-b^{2}}, \\
R^{r\phi r\phi } &=&\frac{-\Lambda \left( \Lambda r^{2}+M\right) }{r^{2}},
\end{eqnarray*}%
and so there is a singularity located at $r=0$ whose horizon is
\begin{equation}
r_{+}=\left( \frac{M+\sqrt{M^{2}+4\Lambda b^{2}}}{-2\Lambda }\right) ^{1/2}.
\label{rp1}
\end{equation}%
Applying Dolgov-Kawasaki stability method, one obtains

\begin{equation}
F_{RR}=\frac{C}{\left( 4\Lambda \right) ^{2}}e^{\frac{R}{4\Lambda }},
\label{ddFrot}
\end{equation}%
which confirms that this model can be stable.

\subsubsection{Rotating solution: case II:}

Here, we generalize the recent rotating spacetime to the case of two unknown
metric functions
\begin{equation}
ds^{2}=-g(r)dt^{2}+\frac{dr^{2}}{g(r)}+r^{2}\left( d\phi +\frac{b(r)}{r^{2}}%
dt\right) ^{2}.  \label{Met3}
\end{equation}
Inserting this metric into the field equations (\ref{eqaa}) and (\ref{eqbb}%
), one can obtain
\begin{eqnarray}
g(r) &=&-\Lambda r^{2}-M-\frac{M^{2}}{4\Lambda r^{2}},  \label{g2} \\
b(r) &=&\pm \left( \sqrt{-\Lambda }r^{2}-\frac{M}{2\sqrt{-\Lambda }}\right)
\label{e2}
\end{eqnarray}
where $\Lambda $ and $M$ are integration constants which are related to the
\textit{negative} cosmological constant and mass parameter, respectively. In
addition, there is a constraint on the $f(R)$ model as
\begin{equation}
f_{R}+\frac{f(R)}{4A}-\frac{1}{2}=0,  \label{eq}
\end{equation}
with the following solution
\begin{equation}
f(R)=-2\Lambda +Ce^{\frac{R}{4\Lambda }}.  \label{f(R)2}
\end{equation}
Indeed, the mentioned rotating solution is valid only for the exponential
form of $f(R)$ model. Calculating the nonzero components of the Riemann
tensor shows that there is a singularity at $r=0$ with the following horizon
\begin{equation}
r_{+}=\sqrt{\frac{-M}{2\Lambda }}.  \label{rp2}
\end{equation}
The stability discussion of the mentioned model is the same as the former
rotating solution.

\section{Conclusions}

In this paper, in order to better understanding of $(2+1)$-dimensional
gravity, we have considered $F(R)$ theories of gravity and searched for
either exact or near horizon solutions of general and specific models of $%
F(R)$.

At first, we showed that the general $F(R)$ gravity with constant Ricci
scalar admits the uncharged static BTZ solution as an exact solution.
Besides it, in the case $F(R)=0$, where the action of the gravity vanishes
on-shell, interestingly, there is a charged solution just the same as that
in the Einstein-PMI gravity when the nonlinearity parameter is chosen $s=3/4$%
.

We also focused attention on the near-horizon region by truncating the black
hole metric to its leading terms close to the horizon. We started with a
trivial $F(R)$ model as an example and generalized our method to general
models of $F(R)$ gravity with (non)constant Ricci scalar and found that the
near horizon metric functions are the same as exact uncharged BTZ solutions
with an additional linear term for nonconstant Ricci scalar case.

Furthermore, we considered specific $F(R)$ models to obtain exact solutions.
We showed that one can obtain asymptotically Lifshitz solution with a
hyperscaling overall factor and Lifshitz solution as a vacuum solution. We
should note that in general we cannot obtain the mentioned exact solutions.
For example, considering the Starobinsky model, we could not obtain
asymptotically Lifshitz solution. In this case, we added an $R^{2}$ term
\cite{Dev} to obtain generalized Starobinsky model with (asymptotically)
Lifshitz solution. Finally, we achieved two kinds of rotating solutions with
rotating black hole interpretation. It is notable that in order to have
rotating solutions, $F(R)$ should be in exponential form.

In this paper, we obtained some near horizon and exact solutions of $F(R)$
gravity and stress on the geometric properties of the solutions. Therefore
it is worthwhile to think about thermodynamic properties of the obtained
solutions.

\section{Acknowledgement}

We are grateful to M. H. Vahidinia and H. Mohammadpour for reading the
manuscript. We also wish to thank Shiraz University Research Council. This
work has been supported financially by Research Institute for Astronomy \&
Astrophysics of Maragha (RIAAM), Iran.


\begin{thebibliography}{99}
\bibitem{DE} A. G. Riess, et al.: Astron. J. \textbf{116}, 1009 (1998).

\bibitem{ModGrav} S. Capozziello, Int. J. Mod. Phys. D \textbf{11}, 483
(2002);

C. Eling, R. Guedens and T. Jacobson, Phys. Rev. Lett. \textbf{96}, 121301
(2006);

E. Elizalde and P. J. Silva, Phys. Rev. D \textbf{78}, 061501 (2008);

R. Brustein and M. Hadad, Phys. Rev. Lett. \textbf{103}, 101301 (2009).

\bibitem{HendiGRG} S. H. Hendi, B. Eslam Panah and S. M. Mousavi, Gen.
Relativ. Gravit. \textbf{44}, 835 (2012).

\bibitem{Early} A. A. Starobinsky, Phys. Lett. B \textbf{91}, 99 (1980);

K. Bamba and S. D. Odintsov, JCAP \textbf{04}, 024 (2008);

\bibitem{Late} S. M. Carroll, etal., Phys. Rev. D \textbf{70}, 043528 (2004);

A. de la Cruz-Dombriz and A. Dobado, Phys. Rev. D \textbf{74}, 087501 (2006);

S. Fay, R. Tavakol and S. Tsujikawa, Phys. Rev. D \textbf{75}, 063509 (2007).

\bibitem{hierarchy} G. Cognola, etal., Phys. Rev. D \textbf{73}, 084007
(2006).

\bibitem{singularity} M. C. B. Abdalla, S. Nojiri and S. D. Odintsov, Class.
Quantum Gravit. \textbf{22}, L35 (2005);

F. Briscese, E. Elizalde, S. Nojiri and S. D. Odintsov, Phys. Lett. B
\textbf{646}, 105 (2007);

S. Nojiri and S. D. Odintsov, Phys. Rev. D \textbf{78}, 046006 (2008);

K. Bamba, S. Nojiri and S. D. Odintsov, JCAP \textbf{10}, 045 (2008);

T. Kobayashi and K. Maeda, Phys. Rev. D \textbf{79}, 024009 (2009).

\bibitem{F(R)DE} S. Capozziello, V. F. Cardone and A. Troisi, Phys. Rev. D
\textbf{71}, 043503 (2005);

S. Nojiri and S. D. Odintsov, Phys. Rev. D \textbf{74}, 086005 (2006);

C. G. Boehmer, T. Harko and F. S. N. Lobo, Astropart. Phys. \textbf{29}, 386
(2008).

\bibitem{F(R)Max} Z. Xue and S. L. Cui, [arXiv:1306.2082].

\bibitem{F(R)nonMax} D. Momeni, M. Raza and R. Myrzakulov, Eur. Phys. J.
Plus \textbf{129}, 30 (2014).

\bibitem{ghosts} A. D. Dolgov and M. Kawasaki, Phys. Lett. B \textbf{573}, 1
(2003);

S. Nojiri and S. D. Odintsov, Phys. Rev. D \textbf{77}, 026007 (2008);

V. Faraoni, Phys. Rev. D \textbf{74}, 104017 (2006);

I. Sawicki and W. Hu, Phys. Rev. D \textbf{75}, 127502 (2007).

\bibitem{Nojiri-Odintsov} S. Nojiri and S. D. Odintsov, Phys. Lett. B
\textbf{657}, 238 (2007);

S. Nojiri and S. D. Odintsov, Phys. Lett. B \textbf{652}, 343 (2007).

\bibitem{F(R)1} M. Akbar and R. G. Cai, Phys. Lett. B \textbf{635}, 7 (2006);

M. Akbar and R. G. Cai, Phys. Lett. B \textbf{648}, 243 (2007);

C. Corda, Astropart. Phys. \textbf{34}, 587 (2011);

C. Corda, Gen. Relativ. Gravit. \textbf{40}, 2201 (2008);

G. Cognola, etal., Phys. Rev. D \textbf{77}, 046009 (2008);

T. P. Sotiriou and V. Faraoni, Rev. Mod. Phys. \textbf{82}, 451 (2010);

S. H. Hendi, Phys. Lett. B \textbf{690}, 220 (2010);

A. De Felice and S. Tsujikawa, Living Rev. Rel. \textbf{13}, 3 (2010).

\bibitem{Hu-Sawicki} W. Hu and I. Sawicki, Phys. Rev. D \textbf{76}, 064004
(2007).

\bibitem{Starobinsky} A. A. Starobinsky, JETP Lett. \textbf{86}, 157 (2007).

\bibitem{Dev} A. Dev, etal., Phys. Rev. D \textbf{78}, 083515 (2008).

\bibitem{Appleby-Battye} S. A. Appleby and R. A. Battye, Phys. Lett. B
\textbf{654}, 7 (2007).

\bibitem{Tsujikawa} S. Tsujikawa, Phys. Rev. D \textbf{77}, 023507 (2008).

\bibitem{BTZnew} L. Hodgkinson and J. Louko, Phys. Rev. D \textbf{86},
064031 (2012);

T. Moon and Y. S. Myung, Phys. Rev. D \textbf{86}, 124042 (2012);

M. Eune, W. Kim and S. H. Yi, JHEP \textbf{03}, 020 (2013);

S. H. Hendi, JHEP \textbf{03}, 065 (2012);

E. Frodden, M. Geiller, K. Noui and A. Perez, JHEP \textbf{05}, 139 (2013).

\bibitem{BTZ1} S. Carlip, Class. Quantum Gravit. \textbf{12}, 2853 (1995);

A. Ashtekar, J. Wisniewski and O. Dreyer, Adv. Theor. Math. Phys. \textbf{6}%
, 507 (2002);

T. Sarkar, G. Sengupta and B. N. Tiwari, JHEP \textbf{11}, 015 (2006).

\bibitem{BTZ2} E. Witten, Adv. Theor. Math. Phys. \textbf{2}, 505 (1998);

S. Carlip, Class. Quantum Gravit. \textbf{22}, R85 (2005).

\bibitem{BTZ3} M. Banados, C. Teitelboim and J. Zanelli, Phys. Rev. Lett.
\textbf{69}, 1849 (1992);

M. Banados, M. Henneaux, C. Teitelboim and J. Zanelli, Phys. Rev. D \textbf{%
48}, 1506 (1993);

S. Nojiri and S. D. Odintsov, Mod. Phys. Lett. A \textbf{13}, 2695 (1998);

R. Emparan, G. T. Horowitz and R. C. Myers, JHEP \textbf{01}, 021 (2000);

S. Hemming, E. K. Vakkuri and P. Kraus, JHEP \textbf{10}, 006 (2002);

M. R. Setare, Class. Quantum Gravit. \textbf{21}, 1453 (2004);

B. Sahoo and A. Sen, JHEP \textbf{07}, 008 (2006);

M. R. Setare, Eur. Phys. J. C \textbf{49}, 865 (2007);

M. Cadoni and M. R. Setare, JHEP \textbf{07}, 131 (2008);

M. I. Park, Phys. Rev. D \textbf{77}, 026011 (2008);

M. I. Park, Phys. Rev. D \textbf{77}, 126012 (2008);

J. Parsons and S. F. Ross, JHEP \textbf{04}, 134 (2009);

M. R. Setare and M. Jamil, Phys. Lett. B \textbf{681}, 469 (2009).

\bibitem{BTZ4} E. Witten, [arXiv:0706.3359].

\bibitem{BTZ6} M. Cadoni and M. R. Setare, JHEP \textbf{07}, 131 (2008);

L. Claessens, [arXiv:0912.2245];

J. de Boer, M. M. Sheikh-Jabbari and J. Simon, Class. Quantum Gravit.
\textbf{28}, 175012 (2011).

\bibitem{BTZ5} S. Hyun, J. Korean Phys. Soc. \textbf{33}, S532 (1998);

F. Canfora and A. Giacomini, Phys. Rev. D \textbf{82}, 024022 (2010);

S. H. Hendi, Eur. Phys. J. C \textbf{71}, 1551 (2011).

\bibitem{Dombriz} A. de la Cruz-Dombriz, A. Dobado and A. L. Maroto, Phys.
Rev. D \textbf{80}, 124011 (2009).

\bibitem{HendinonlinearEPJC} S. H. Hendi, Eur. Phys. J. C \textbf{69}, 281
(2010).

\bibitem{foot} One may see the Wald entropy $S=\frac{A}{4\pi}(1+f_{R})$
vanishes for this theory.

\bibitem{near} J. W. York, Phys. Rev. D \textbf{28}, 2929 (1983);

W. H. Zurek and K. S. Thorne, Phys. Rev. Lett. \textbf{54}, 2171 (1985);

J. A. Wheeler, \emph{A Journey into Gravity and Spacetime Freeman} (N.Y.),
(1990);

G. 't Hooft, Nucl. Phys. B \textbf{335}, 138 (1990);

V. Frolov and I. Novikov, Phys. Rev. D \textbf{48}, 4545 (1993);

S. Carlip, Phys. Rev. D \textbf{51}, 632 (1995);

M. Cvetic and A. Tseytlin, Phys. Rev. D \textbf{53}, 5619 (1996);

A. Strominger, JHEP \textbf{02}, 009 (1998).

\bibitem{HendiMomeni} S. H. Hendi and D. Momeni, Eur. Phys. J. C \textbf{71}%
, 1823 (2011).

\bibitem{DKstability1} A. D. Dolgov and M. Kawasaki, Phys. Lett. B \textbf{%
573}, 1 (2003);

S. Nojiri and S. D. Odintsov, Phys. Rev. D \textbf{68}, 123512 (2003).

\bibitem{DKstability2} V. Faraoni, Phys. Rev. D \textbf{74}, 104017 (2006);

V. Faraoni, Phys. Rev. D \textbf{76}, 127501 (2007);

O. Bertolami and M. C. Sequeira, Phys. Rev. D \textbf{79}, 104010 (2009).
\end{thebibliography}
\end{document}